\documentstyle[12pt]{article}
\let\oldsection=\section
\def\section#1{\oldsection{#1}\indent}
\def\GeV{{\rm\: GeV}}

\title{ \Large{
The magnitude of the gluon condensate and the masses of $c$ and
$b$  quarks from the families of ${\cal J}$/$\Psi$ and $\Upsilon$ mesons.}  }

\author{
    B.V.Geshkenbein\footnotemark[1] \\
  \small{Institute for Theoretical and Experimental Physics} \\
  \small{SU-117279  Moscow  Russia} \\
\\
    V.L.Morgunov \footnotemark[2]\\
  \small{Institute for Theoretical and Experimental Physics\footnotemark[3]}\\
  \small{SU-117279  Moscow  Russia} \\
  \small{and}\\
  \small{Physics Research Division\footnotemark[4]}\\
  \small{Superconducting Super Collider Laboratory}\\
  \small{2550 Beckleymeade Avenue}\\
  \small{Dallas, Texas  USA}\\
        }
\date{February 10, 1994}

\begin{document}
\vbox{%
\maketitle}

\begin{abstract}

It is shown that the model used in [1] for the description  of the
experimental function $R_c (s)$ in the form of a sum of $\delta$-functions
and the plateau contradicts the Wilson operator expansion (O.E.) in
the terms due to the gluon condensate. A QCD model with an
infinite number of vector mesons [12,13] does satisfy the
requirements of the O.E. for masses and electronic widths resonances
close to the experimental values. The region of allowable values of the
masses of $c$ and $b$-quarks and the gluon condensate compatible with the O.E.
is
obtained (Figs. 2,3).
\end{abstract}

\section{Introduction}

\footnotetext[1]{E-mail : geshken@vxitep.itep.msk.su}
\footnotetext[2]{E-mail : morgunov@vxdesy.desy.de, morgunov@vxcern.cern.ch}
\footnotetext[3]{Permanent address.}
\footnotetext[4]{Operated by the Universities Research Association, Inc.,
for the U.S. Departament of Energy under Contract No. DE-AC35-89ER40486.}

The gluon condensate $ \langle (\alpha_{s}/\pi)G^{2}\rangle $ is one of the
fundamental characteristics of the QCD vacuum. In a
basic paper [1],  Shifman, Vainshtein and Zakharow showed
that the magnitude of the gluon condensate is nonzero and obtained
for it the value
\begin{equation}
     {\langle (\alpha_{s}/\pi)G^{2} \rangle }_{SVZ} = 0.012 \GeV^{4}
\end{equation}
Since then, a large number of papers [2-11] gave various gluon
condensate values which significantly differ from the value (1).

The analysis made for the families of ${\cal J}$/$\Psi$ and $\Upsilon$
mesons in paper [12] gave for the gluon condensate the value
\begin{equation}
    0.05 \leq  \langle (\alpha_{s}/\pi)G^{2} \rangle \leq 0.1 \GeV^{4},
\end{equation}
rather different from value (1). This paper is devoted to determining which
of the values of the gluon condensate is correct.

     The plan of the article is following:

In section 2 it will be shown the model used in [1] for the description
of the experimental function $R_c (s)$  in the form of the sum of
$\delta$-functions and the plateau strongly contradicts the Wilson
operator expansion (O.E.) in terms due to the gluon condensate.

In section 3 we will develop a QCD model with an infinite number of vector
mesons [12,13]. This model allows us to satisfy the requirements of the O.E.
It will be shown that the basic equation (20) can be derived without the added
assumption on $k \gg 1$.

In section 4 we will perform numerical calculations for both meson families,
where we will obtain the region of the allowable values of the gluon
condensate and the masses of the $c$ and $b$ quarks consistent with the O.E.

\section{What is the reason of our doubts of the
   validity of the gluon condensate value (1)?}

The O.E. lies at the basis of the calculations of the gluon condensate value.
The O.E. for the T-product of vector currents composed of $c$
quarks has the form [1]
\begin{equation}
     i \int d^4 x e^{iqx} T \left\{ \jmath^{c}_{\mu}(x),
     \jmath^{c}_{\nu}(0)\right\}
      = (q_{\mu}q_{\nu}-q^{2}g_{\mu\nu}) \ (C_I I + C_G O_G + \cdots)
\end{equation}
where $ \jmath^{c}_{\mu}(x) = \overline{c} \  (x)  \gamma_{\mu} \ c(x)$,
$I$ is the unit operator and
\begin{equation}
  C_{I}=\Pi^{c}_{I}(Q^2)= \frac{1}{12\pi^{2}Q_{c}^{2}}
  \int\limits_{4m_{c}^{2}}^{\infty} \frac{R_{c}^{PT}(s)\/{\em{ds}}}{s+Q^{2}}
\end{equation}
where  $Q^2 = -q^2$ ,  $Q_c$ is a $c$ quark charge, and
$C_{I}$ is the coefficient function of the unit operator.

The function $R^{PT}_{c}$ includes all corrections on $\alpha_{s}$  in
perturbation
theory (PT). The equation (4) has been written without the subtraction, because
it uses
derivatives of (4) with respect to $Q^2$ only. The remaining values are defined
as: $O_G
= G^{a}_{\mu\nu}(0)G^{a}_{\mu\nu}(0)$,  where $G^{a}_{\mu\nu}$ is the operator
of
the gluon field. \[ C_G=\frac{\alpha_{s}}{\pi}P(Q^2) , \]

\begin{equation}
 P(Q^2)=\frac{1}{48Q^4}
    \left\{ {3(a+1)(1-\frac{1}{a})}^{2}
        \frac{1}{2\sqrt{a}} ln\frac{\sqrt{a}+1}{\sqrt{a}-1}
        - 3+\frac{2}{a}-\frac{3}{a^2} \right\},
\end{equation}

\[ a=1+\frac{4m_{c}^{2}}{Q^2} \]

The O.E. is supposed to be valid for all positive $Q^2$.
If we take the vacuum expectation value of (3) we obtain the
theoretical definition of the polarization operator following from the O.E.
\begin{equation}
	\Pi^{c}_{Theor.}(Q^2)=\Pi^{c}_{I}(Q^2)
	+ \langle (\alpha_{s}/\pi)G^{2} \rangle P(Q^2), Q^2 \geq 0
\end{equation}
The dispersion relation for $\Pi^{c}_{Exp.}(Q^2)$ permits expression of
$\Pi^{c}_{Exp.} (Q^2)$ via the measuring value $R_{c}(s)$
\begin{equation}
   \Pi^{c}_{Exp.}(Q^2)= \frac{1}{12\pi^{2}Q_{c}^{2}}
  \int\limits_{4m_{c}^{2}}^{\infty} \frac{R_{c}(s){\em{ds}}}{s+Q^{2}}
\end{equation}
In the model of [1] it was
\begin{equation}
{(\Pi^{c}_{Exp.}(Q^{2}))}_{SVZ} =
 \frac{1}{12\pi^{2}Q_{c}^{2}}
   \left\{
 \frac{9\pi}{\alpha^{2}}
\sum_{i=0}^{5} \frac{\Gamma_{i}^{ee} M_{i}}{M_{i}^{2}+Q^{2}} +
  \int\limits_{\overline{s}}^{\infty} \frac{R_{c}^{(1)}(s){\em{ds}}}{s+Q^{2}}
    \right\}
\end{equation}
where $\Gamma_{i}^{ee}$ is the electronic width of the i-resonance, $M_i$ is
the
mass of this resonance, $\alpha=1/137$. The summation is over all 6
resonances of ${\cal J}$/$\Psi$ family. In Ref.[1] corrections of the first
order in $\alpha_s$ are taken into account only because the
function $R_{c}^{(1)}(s)$ is chosen in the form
\begin{equation}
  R_{c}^{(1)}(s) = R_{c}^{(0)}(s)
\left\{
   1 + \frac{\alpha_{s}}{3}
\left[
  \frac{2 \pi}{v} - ( \frac{\pi}{2} - \frac{3}{4 \pi} ) (3+v)
\right]
\right\}
\end{equation}
where
\begin{equation}
  R_{c}^{(0)}(s) = \frac{3}{2}Q_{c}^{2} v (3-v^{2})
\end{equation}

\[Q_{c}=\frac{2}{3}, v={(1-\frac{4m_{c}^{2}}{s})}^{1/2}, \alpha_{s}=0.2 \]
For $s > \overline{s}$ $\ R_{c}^{(1)}$ is taken in the simplified form

\begin{equation}
  R_{c}^{(1)}(s) = \frac{4}{3}
(1+\frac{\alpha_{s}}{\pi}) \Theta (s-\overline{s})
\end{equation}

The charmed quark mass $m_{c}=1.26 \GeV$, the  value
$ \overline{s}={(4.2)}^{2}  {\GeV}^{2} $
and the gluon condensate (1) were found from a requirement of minimal
discrepancy
between the first ten moments of ${(\Pi^{c}_{Exp.}(Q^{2}))}_{SVZ}$
and $(\Pi^{c}_{Theor.}(Q^{2}))$.
It can be shown that the formula (8) strongly contradicts the O.E. in
the terms due to the gluon condensate. After subtracting from
formulas (6) and (8) the contribution of the unit operator (4),
we obtain:

\begin{equation}
\Delta \Pi^{c}_{Theor.}(Q^{2})=\langle (\alpha_{s}/\pi)G^{2}\rangle P(Q^{2})
\end{equation}

\begin{equation}
{(\Delta \Pi^{c}_{Exp.}(Q^{2}))}_{SVZ} =
 \frac{1}{12\pi^{2}Q_{c}^{2}}
   \left\{
 \frac{9\pi}{\alpha^{2}}
\sum_{i=0}^{5} \frac{\Gamma_{i}^{ee} M_{i}}{M_{i}^{2}+Q^{2}} -
  \int\limits^{\overline{s}}_{4m_{c}^{2}}
    \frac{R_{c}^{(1)}(s){\em{ds}}}{s+Q^{2}}
    \right\}
\end{equation}

Formula (12) contains only the contribution of  the  operator
connected with the gluon condensate but formula (13) contains in
addition higher dimensional operators. The contribution
of operators of higher dimension is small at
$Q^{2} \geq 0$ so expressions (12) and (13) must be close.
The formulas (12) and (13) have different asymptotes if
$m_{c}=1.26 \GeV, \overline{s}=4.2 {\GeV}^{2}$ (see [1])
since formula (13) contains terms in  $1/Q^{2}$.
The comparison between terms in $1/Q^{4}$ in formulae (12) and (13)
gives
\begin{equation}
{\langle (\alpha_{s}/\pi)G^{2}\rangle } =
 \frac{1}{\pi^{2}Q_{c}^{2}}
   \left\{
 \frac{9\pi}{\alpha^{2}}
\sum_{i=0}^{5} \Gamma_{i}^{ee} M_{i}^{3} -
  \int\limits^{\overline{s}}_{4m_{c}^{2}}
    R_{c}^{(1)}(s) s {\em{ds}}
    \right\}
\end{equation}
Substitution of $m_{c}$ and $\overline{s}$ from  [1] in formula (11)
gives  $\langle (\alpha_{s}/\pi)G^{2}\rangle =-3.85  \GeV^{4}$.
We shall attempt to  improve the situation by changing the parameters
in formula (13). Absence of $1/Q^{2}$ terms in asymptote (12) gives
\begin{equation}
 \frac{9\pi}{\alpha^{2}}
\sum_{i=0}^{5} \Gamma_{i}^{ee} M_{i} -
  \int\limits^{\overline{s}}_{4m_{c}^{2}}
    R_{c}^{(1)} (s) {\em{ds}} = 0
\end{equation}
After the integral in (15) has been taken, the parameters  $m_{c},\overline{s}$
can be expressed in  terms  of $\overline{v}={(1-{4m_{c}^{2}}/
{\overline{s}})}^{1/2},M_{i},{\Gamma}_{i}^{ee}$.  The
parameter $\overline{v}$, the masses $M_{i}$ and the
widths ${\Gamma}_{i}^{ee}$ were varied in
order to make (12) and (13) maximally close.
The masses $M_{i}$ and the widths $\Gamma_{i}^{ee}$ were varied within
the limits of four experimental errors.
 It  is the best result possible  to achieve in the vicinity of (12) and (13)
(its difference is less then 20\%) at values
$Q^{2}  > 4 \GeV^{2}$ only. At $Q^{2} = 0$ the right-hand sides
of (12) and (13) differ by a factor of five.
In our opinion it is too strong a violation of the O.E.. In the next section
a model meeting O.E. requirements  will be constructed.

\section{The QCD model with an infinite number of vector mesons.}

In this section we intend to improve the QCD model with an infinite
number of vector mesons suggested in [12,13]. We  shall  describe
the families ${\cal J}$/$\Psi$ and $\Upsilon$ mesons in the same  way.  Let us
consider  the polarization operator $ \Pi^{a} (Q^{2})$
$a=c$ for the ${\cal J}$/$\Psi$ family, $a=b$ for the $\Upsilon$ family.

The polarization operator  $ \Pi^{a} (Q^{2})$  has the form [12]
\begin{equation}
\Pi^{a}(Q^{2}) =
 \frac{1}{12\pi^{2}Q_{a}^{2}}
 \frac{9\pi}{\alpha^{2}}
\sum_{k=0}^{\infty} \frac{\Gamma_{k}^{ee} M_{k}}{s_{k}+Q^{2}}
\end{equation}
in the approximation  of an infinite number of narrow resonances, having
masses $M_{k}$ and electronic widths $\Gamma_{k}^{ee}$,
$s_{k}=M_{k}^{2}$ ; $Q_a$ is a quark charge.

If the widths and the masses of the vector  mesons obey the conditions
\begin{equation}
    M_{k}^{2} - M_{k-1}^{2} \ll M_{k} \Gamma_{k} \ll M_{k}^{2}
\end{equation}
for $k \geq 5$ ($\Gamma_{k}$ is the total width of the $k$-$th$ resonance),
then after the sixth resonance the function $R(s)$ will be described by a
smooth
curve and all the formulas of the model under consideration  will
be applied as was shown in [13].

The polarization operator (16) can be rearranged
in form with the separated unit operator.
The  remainder  of  the
polarization operator can be associated with the gluon condensate
and with the contribution of  higher  dimensional  operators.

We transform the sum in (13) into an integral  by  means  of  the
Euler-Maclaurin formula beginning from $k = k_0$.
where $k_{0}$ is the middle resonance
number from 6 discovered resonances $(k_{0}=2$ or $k_{0}=3)$.
For definiteness we have chosen $k_{0}=2$. Then we obtain:
\begin{eqnarray}
\Pi^{a} (Q^{2}) =
 \frac{1}{12\pi^{2}Q_{a}^{2}}
 \frac{9\pi}{\alpha^{2}}
   \left\{
  \int\limits^{\infty}_{4m_{a}^{2}}
     \frac{\Gamma_{k}^{ee}M_{k}}{s_{k}+Q^{2}}
     \frac{{\em{dk}}}{{\em{ds_{k}}}} {\em{ds_{k}}} -
  \int\limits^{s_{2}}_{4m_{a}^{2}}
     \frac{\Gamma_{k}^{ee}M_{k}}{s_{k}+Q^{2}}
     \frac{{\em{dk}}}{{\em{ds_{k}}}} {\em{ds_{k}}} + \right. \nonumber   \\
   \left.
    \frac{\Gamma_{0}^{ee} M_{0}}{s_{0}+Q^{2}}+
    \frac{\Gamma_{1}^{ee} M_{1}}{s_{1}+Q^{2}} +
    \frac{1}{2} \frac{\Gamma_{2}^{ee} M_{2}}{s_{2}+Q^{2}}-
    \frac{1}{12} \frac{{\em{d}}}{{\em{dk}}}
 ( \frac{\Gamma_{k}^{ee} M_{k}}{s_{k}+Q^{2}}){|}_{k=2} + \right. \nonumber   \\
   \left.
    \frac{1}{720} \frac{{\em{d^{3}}}}{{\em{dk^{3}}}}
  (\frac
{\Gamma_{k}^{ee} M_{k}}
 {s_{k}+Q^{2}}
){|}_{k=2} - \cdots \right\}
\end{eqnarray}
It seems reasonable to set the first term on the right-hand side of
(18) equal to the first term on the right-hand side of (6)
\begin{equation}
 \frac{1}
{12\pi^{2}Q_{a}^{2}}
 \frac{9\pi}{\alpha^{2}}
   \int\limits^{\infty}_{4m_{a}^{2}}
     \frac{\Gamma_{k}^{ee}M_{k}}{s_{k}+Q^{2}}
     \frac{{\em{dk}}}{{\em{ds_{k}}}} {\em{ds_{k}}} =
 \frac{1}{12\pi^{2}Q_{a}^{2}}
\int\limits^{\infty}_{4m_{a}^{2}}
\frac{R_{a}^{PT}(s){\em{ds}} }{s+Q^{2}}
\end{equation}

We consider the equality (19) as an ansatz allowing separation of the large
terms associated with the unit operator from small terms associated with
the operator connected with the gluon condensate.
(19) has one and only one solution
\begin{equation}
\Gamma_{k}^{ee} =
\frac{2\alpha^{2}}{9\pi}
R_{a}^{PT}(s_{k}) M_{k}^{(1)}
\end{equation}
In the equation (20) and in the following formulas we use the notation
\begin{equation}
M_{k}^{(l)} \equiv \frac {{\em{d}}^{l}M_{k}}{{\em{d}}k^{l}},
s_{k}^{(l)} \equiv \frac {{\em{d}}^{l}s_{k}}{{\em{d}}k^{l}}
\end{equation}

It is significant that the derivation of formula (20)  does not demand the
additional assumption $k \gg 1$ as in [13].
It may be proven that the function $s_{k} \equiv s(k)$ given at
$k = 0, 1, 2, \ldots $ can be continued by an analytical function of the
complex variable $k$ with the cut along the negative axes. We will
not use the analytical properties of the function $s(k)$; it is suggested
that the function $s(k)$ is continuous, differentiable with respect to
$k$ and has the necessary number of (not too great) derivatives
with respect to $k$ at $k = 2$. The derivatives $s_{2}^{(l)}$ will
be considered as parameters.

In these notations the contribution of  terms  connected  with
the gluon condensate and higher dimensional operators is
\begin{eqnarray}
\Delta \Pi^{a}(Q^{2}) =
 \frac{1}{12\pi^{2}Q_{a}^{2}}
   \left\{
 \frac{R_{a}^{PT}(s_{0})s_{0}^{(1)}}{s_{0}+Q^{2}}+
 \frac{R_{a}^{PT}(s_{1})}{s_{1}+Q^{2}}-
  \int\limits^{s_{2}}_{4m_{a}^{2}}
    \frac{R_{a}^{PT}(s){\em{ds}}}{s+Q^{2}}+  \right. \nonumber   \\
   \left.
    \frac{1}{2} W_{PT}(Q^{2}) - \frac{1}{12} W_{PT}^{(1)}(Q^{2}) +
    \frac{1}{720} W_{PT}^{(3)}(Q^{2}) - \cdots
    \right\}
\end{eqnarray}
Here we have introduced the notations:
\begin{equation}
W_{PT}(Q^{2}) = \frac{R_{a}^{PT}(s_{2})s_{2}^{(1)}}{s_{2}+Q^{2}},
W_{PT}^{(l)}(Q^{2}) = \frac{{\em{d^{l}}}}{{\em{dk^{l}}}} \left(
\frac{R_{a}^{PT}(s_{k})s_{k}^{(1)}}{s_{k}+Q^{2}} \right) {|}_{k=2}
\end{equation}
Because it is suggested that the contribution of higher  dimensional
operators is small we obtain:
\begin{equation}
\Delta \Pi^{a}(Q^{2}) = \langle (\alpha_{s}/\pi)G^{2}\rangle P(Q^{2})
\end{equation}
The function $P(Q^2)$ is defined by formula (5) for $Q^2 > 0$. For
$-4 m^{2}_{a} < Q^2 < 0$ the function  $P(Q^2)$  is obtained
by analytic continuation[12].

\begin{equation}
 P(Q^2)=\frac{1}{48Q^4}
    \left\{\frac{3(a+1) (a-1)^{2}}{a^2} R
        - 3+\frac{2}{a}-\frac{3}{a^2} \right\}
\end{equation}

\[{ R = - \frac{1}{\sqrt{-a}}  \arctan \frac{1}{\sqrt{-a}} }\]

\[ a=1+\frac{4m_{c}^{2}}{Q^2} \]

The O.E. is suggested to be valid for the family of  ${\cal J}$/$\Psi$ mesons
for $Q^2 > 0$ and for the family of $\Upsilon$ mesons for
$Q^2 > -4m_{b}^{2}+10 \GeV^2$
The function $P(Q^2)$ for $c$ and $b$ quarks is shown in Figure 1.
The reasonableness in choosing the region of validity of the O.E.
for the families of  ${\cal J}$/$\Psi$ and  $\Upsilon$ mesons can be seen from
Figure 1.

\begin{figure}
   \vspace{7cm}
   \caption{The function $P(Q^{2})$ for $c$ and $b$ quarks.}
\end{figure}

On the right-hand side of (24) corrections in $\alpha{_s}$ are not taken
into account and consequently, on the left-hand side, too, these
corrections can be disregarded, i.e., we can make the  replacement
$R_{a}^{PT}{(s)} \rightarrow R_{a}^{0}{(s)} = (3/2)Q_{a}^{2} v(3-
v^{2})$\footnote{It is supposed that the masses of the resonances considered
do not depend on $\alpha_{s}$. As an illustration we refer to [14], where
is shown that the mass difference between a proton and a  neutron
may not be electromagnetic but has its origin in the  difference of the
masses of $u$ and $d$-quarks.}
 As a result we write the left-hand side of (22) in the form
\begin{eqnarray}
 \Delta \Pi^{a}(Q^{2}) = \frac{1}{8\pi^{2}} \left\{
W_{0} + W_{1} - 2 \frac{4m_{a}^{2}}{Q^{2}} F + \frac{1}{2}W_{2} -
 \frac{1}{12}W_{2}^{(1)} +\frac{1}{720}W_{2}^{(3)} - \right. \nonumber \\
\left.
-\frac{1}{30240}W_{2}^{(5)} + \frac{1}{129600}W_{2}^{(7)}
- \frac{1}{47900160}W_{2}^{(9)} + \cdots \right\}
\end{eqnarray}
In (26) we use the notations
\[ W_{k} = \frac{u_{k}}{s_{k}+Q^{2}},
   u_{k}=v_{k}(3-v_{k}^{2})s_{k}^{(1)}, k=0,1,2 \ldots \]

\begin{equation}
F=\frac{Q^{2}}{8m_{a}^{2}}
  \int\limits_{4m_{a}^{2}}^{s_{2}} \frac{v(3-v^{2}){\em{ds}}}{s+Q^{2}}=
\frac{1}{a-1} \left[
ln\frac{1+v_{2}}{1-v_{2}} - \frac{(3-a)\sqrt{a}}{2}
ln \frac{\sqrt{a}+v_2{}}{\sqrt{a}-v_2{}} \right] - v_2
\end{equation}

\[ a=1+ \frac{4m_{a}^{2}}{Q^{2}} \]
For $-4m_{a}^{2}  <  Q^2  < 0$ the following formula holds for the function F:
\begin{equation}
F=\frac{1}{a-1} \left[
ln\frac{1+v_{2}}{1-v_{2}} - (3-a)\sqrt{-a} \;
arctg \frac{v_2}{\sqrt{-a}} \right]
\end{equation}
The superscripts on $W_{2}$ are explained by (21).
For $Q^{2} \rightarrow \infty$ the right-hand side  of (24) takes the form
\begin{equation}
\langle (\alpha_{s}/\pi)G^{2}\rangle P(Q^{2}) \rightarrow
 -\frac{1}{12} \langle (\alpha_{s}/\pi)G^{2}\rangle \frac{1}{Q^{4}}
\end{equation}
Consequently, the terms $1/Q^{2}$ on the  left-hand  side  of (21)
should cancel, and we obtain the equation:
\begin{eqnarray}
C_{2} = u_{0}+u_{1}-2s_{2}v_{2}^{3} +
   \frac{1}{2}u_{2} -
   \frac{1}{12}u_{2}^{(1)} +
   \frac{1}{720}u_{2}^{(3)} -
   \frac{1}{30240}u_{2}^{(5)} + \nonumber \\
 + \frac{1}{1209600}u_{2}^{(7)} -
   \frac{1}{47900160}u_{2}^{(9)}+\cdots = 0
\end{eqnarray}
Equating the terms $1/Q^{4}$ as $Q^{2} \rightarrow \infty$  on both sides
of (24),  we obtain:
\begin{eqnarray}
{\langle (\alpha_{s}/\pi)G^{2}\rangle}_{\infty}= \frac{3}{2\pi^{2}} \left\{
  u_{0}s_{0}+u_{1}s_{1}-2{(4m_{a}^{2})}^{2}
F_{4}+\frac{1}{2} u_{2}s_{2} -
  \frac{1}{12} {(u_{2}s_{2})}^{(1)} +  \right. \nonumber \\
\left.
 + \frac{1}{720} {(u_{2}s_{2})}^{(3)} -
   \frac{1}{30240} {(u_{2}s_{2})}^{(5)} +
   \frac{1}{1209600} {(u_{2}s_{2})}^{(7)} - \right. \nonumber \\
\left.
 - \frac{1}{47900160} {(u_{2}s_{2})}^{(9)}+\cdots \right\}
\end{eqnarray}

\begin{equation}
F_{4} = \frac{1}{2{(4m_{a}^{2})}^{2}}
  \int\limits_{4m_{c}^{2}}^{s_{2}} sv(3-v^{2}){\em{ds}}=
          \frac{v_{2}(3-v_{2}^{2})}{4(1-v_{2}^{2})^2} -
          \frac{3}{8} ln\frac{1+v_{2}}{1-v_{2}}
\end{equation}
where ${\langle (\alpha_{s}/\pi)G^{2}\rangle }_{\infty} $ is the value
of the gluon  condensate  from (31).
In (22,25,26) the terms not written out have been omitted
because of the smallness of the coefficients in these terms.

    The unobserved parameters $s_{2}^{(k)} (k=2, ...10)$ may be  connected
with the measured quantities $s_{i}, s_{i}^{(1)}(i=0,1,3,4,5)$ by  Taylor
expansion.
\begin{equation}
s_{i} = \sum_{k=0}^{10} \frac{{(i-2)}^{k}}{k!}s_{2}^{(k)}, i=0,1,2,3,4,5
\end{equation}
\begin{equation}
s_{i}^{(1)} = \sum_{k=0}^{9} \frac{{(i-2)}^{k}}{k!}s_{2}^{(k+1)}, i=0,1,2,3,4,5
\end{equation}
The Taylor expansions are cut off at k=10 so that (33-34) contain
the same parameters $s_{2}^{(2)},s_{2}^{(3)}, $ $\ldots, s_{2}^{(10)}$
as (26,30,31).

     The electronic widths of the resonances $\Gamma_{k}^{ee}$ are given by
(20). For
$R_{a}^{PT} (s)$ we shall  use  the
formula
\begin{equation}
R_{a}^{PT} (s_{i}) = R_{c}^{(0)} (s_{i}) {\cal {D}}_{i}(v_{i})
\end{equation}
where $R_{a}^{(0)} (s)$ is given  by  (10)  and  for  function
$D_{k}(v_{k})$ the following expression holds:
\begin{equation}
{\cal {D}}_{i}(v_{i}) =
\frac{4\pi \alpha_{s}/3v_{i}}{1-exp(-4\pi \alpha_{s}/3v_{i})} -
\frac{1}{3}(\frac{\pi}{2} -\frac{3}{4\pi}) (3+v_{i}) \alpha_{s}
\end{equation}
In (35) we have taken into account terms of the first  order  in
$\alpha_{s}$ and Coulomb terms of all orders in $\alpha_{s}/v_{k}$.
The Coulomb  poles are taken into account in the same way as in [12].

Using (16,25,28,31) the unobserved parameters
$s_{2}^{(k)} (k=2, ...10)$ as well as the electronic widths
$\Gamma_{0}^{ee}$ and $\Gamma_{1}^{ee}$ are expressed by the
masses  $M_{i}  (i=0,  ...5)$  and  the
electronic  widths  $\Gamma_{i}^{ee}(i=2,  ...5)$.

Note that the order of the gluon condensate magnitude is $ \langle
(\alpha_{s}/\pi)G^{2}\rangle \sim s_{0} s_{0}^{(1)} $
from dimensionality (31). It is $ \langle (\alpha_{s}/\pi)G^{2}\rangle
\sim 30 \GeV^4 $ for the ${\cal J}$/$\Psi$ family and
$ \langle (\alpha_{s}/\pi)G^{2}\rangle
\sim 1000 \GeV^4 $ for  the $\Upsilon$  family. As obtained in this work, the
value
$ \langle (\alpha_{s}/\pi)G^{2}\rangle \approx 0.05 \div 0.1 \GeV^4 $
from the analysis of  the ${\cal J}$/$\Psi$ and $\Upsilon$  families is due to
great
cancellations in (26,31). Analogous cancellations occur in model
[1]. These cancellations make calculations very difficult. There is a very
great
sensitivity in all results of the values of resonance masses $M_{i} $
 and electronic widths $\Gamma _{i}^{ee}$.
A negligibly small variation of the magnitudes $M_{i} $ and $\Gamma _{i}^{ee}$
gives disagreement in (24).

\section{Numeric calculations and results.}

Let us point out the necessity of fine adjustment at the masses and
the electronic widths of the mesons considered. The right-hand and left-hand
sides
of (24)
differ several times at the experimental values of masses $(M_{i})_{Exp.}
(i = 0,1, ...5)$ and electronic widths $(\Gamma_{i}^{ee})_{Exp.}(i=2,  ...5)$ .
That is why  the masses and the electronic widths of resonances were varied
in addition to varying of quarks masses and gluon condensate.
The mawith the mass of the c-quark $m_{c}= 1.302 \GeV$ ${\langle
(\alpha_{s}/\pi)G^{2}\rangle}
= 0.0526 \GeV^{4}$
(from the ${\cal J}$/$\Psi$  family) and the mass of the b-quark
$m_{b} = 4.547 \GeV$ ${\langle (\alpha_{s}/\pi)G^{2}\rangle} = 0.0615 \GeV^{4}$
(from the $\Upsilon$  family).
At these values right-hand and left-hand sides of (24) coincide with an
accuracy better than $3\%$ for the ${\cal J}$/$\Psi$  family and with accuracy
better than $9\%$ for the $\Upsilon$  family. Thereafter, the masses of the
quarks and the magnitude of the gluon condensate were
varied over the region where the right-hand and left-hand sides of (24)
coincided with an accuracy better than $20\%$; this region is called
the permissible region.
%
%
The permissible region for the c-quark is presented in Figure 2 and for the
b-quark in
Figure 3. The best magnitude of the gluon condensate follows from Figs. 2,3
\begin{equation}
   0.04 \leq {\langle (\alpha_{s}/\pi)G^{2}\rangle} \leq 0.105  \GeV^{4}
\end{equation}

\begin{table}[t]

Table 1. The domain of variations of masses and the electronic widths
of resonances, in GeV.

\begin{center}
${\cal J}$/$\Psi$  family

\begin{tabular}{|l|c|c|c|c|c|c|}
\hline
&0&1&2&3&4&5 \\
\hline
$M_{i,Theor.}^{Best}$
& 3.09688 & 3.68605 & 3.77464 & 4.00219 & 4.08354 & 4.40874  \\
\hline
$M_{i,Theor.}^{max}$
& 3.09790 & 3.68705 & 3.78337 & 4.02603 & 4.14069 & 4.42646  \\
\hline
$M_{i,Theor.}^{min}$
& 3.09576 & 3.68539 & 3.76689 & 4.00013 & 4.08176 & 4.39045  \\
\hline
$M_{i,Exp.}$
& 3.09693 & 3.686   & 3.7699  & 4.04    & 4.159   & 4.415    \\
\hline
$Exp. err.$
& 0.00009 & 0.0001  & 0.0025  & 0.01    & 0.02    & 0.006    \\
\hline
\end{tabular}


\begin{tabular}{|l|c|c|c|c|c|c|}
\hline
&0&1&2&3&4&5 \\
\hline
$\Gamma_{i,Theor.}^{Best}$
& 5.2896 & 1.9379 & 0.31942 & 1.6366 & 0.67065 & 0.51277  \\
\hline
$\Gamma_{i,Theor.}^{max}$
& 6.1576 & 2.0227 & 0.3999  & 1.8489 & 1.4910  & 0.8740   \\
\hline
$\Gamma_{i,Theor.}^{min}$
& 5.0393 & 1.6147  & 0.2593  & 1.4665 & 0.1142 & 0.2006   \\
\hline
$\Gamma_{i,Exp.}$
& 5.36   & 2.14    & 0.26    & 0.75   & 0.77    & 0.47   \\
\hline
$Exp. err.$
& 0.29   & 0.21    & 0.04    & 0.15   & 0.23    & 0.1   \\
\hline
\end{tabular}

$\Upsilon$  family

\begin{tabular}{|l|c|c|c|c|c|c|}
\hline
&0&1&2&3&4&5 \\
\hline
$M_{i,Theor.}^{Best}$
& 9.46032 & 10.0220 & 10.3553 & 10.5841 & 10.8506 & 11.0102 \\
\hline
$M_{i,Theor.}^{max}$
& 9.46074 & 10.0239  & 10.3570  & 10.5870   & 10.8623  & 11.0349 \\
\hline
$M_{i,Theor.}^{min}$
& 9.45986 & 10.0220  & 10.3543  & 10.5736   & 10.8494  & 11.0031 \\
\hline
$M_{i,Exp.}$
& 9.46032 & 10.0233  & 10.3553  & 10.58     & 10.865   & 11.019  \\
\hline
$Exp. err.$
& 0.00022 & 0.00031  & 0.0005   & 0.0035    & 0.008    & 0.008  \\
\hline
\end{tabular}


\begin{tabular}{|l|c|c|c|c|c|c|}
\hline
&0&1&2&3&4&5 \\
\hline
$\Gamma_{i,Theor.}^{Best}$ & 1.0217 & 0.64051 & 0.34043 & 0.35361 & 0.37848 &
0.10799  \\
\hline
$\Gamma_{i,Theor.}^{max}$ & 1.1876 & 0.6452 & 0.4105 & 0.3536 & 0.4500 &
0.1899 \\
\hline
$\Gamma_{i,Theor.}^{min}$ & 1.0198 & 0.5913 & 0.3404 & 0.3174 & 0.3617 &
0.0725  \\
\hline
$\Gamma_{i,Exp.}$ & 1.34     & 0.56      & 0.44      & 0.24       & 0.31      &
0.13     \\
\hline
$Exp. err.$ & 0.04     & 0.09      & 0.04      & 0.05       & 0.07      & 0.03
    \\
\hline
\end{tabular}
\end{center}
\end{table}

\begin{figure}
  \vspace{7cm}
  \caption{The domain of allowable magnitudes of gluon condensate
     and c-quark mass. The difference less than $20\%$ within the curve for all
     $Q^{2} > 0$.}
\end{figure}
\begin{figure}
   \vspace{7cm}
   \caption{The domain of allowable magnitudes of gluon condensate and
    b-quark mass. The difference less than $20\%$ within the curve for all
    $Q^{2}$ from interval $-4 m_{b}^{2} + 10 \GeV^{2} < Q^{2} < \infty$.}
\end{figure}

The magnitude of the gluon condensate obtained from (31),
${\langle (\alpha_{s}/\pi)G^{2}\rangle}_{\infty}$, practically coincides with
${\langle (\alpha_{s}/\pi)G^{2}\rangle}$.
The masses and the electronic widths of the resonances were varied in the
permissible region from $(M_{i}^{ })_{min}$ , $(\Gamma_{i}^{ee})_{min}$ to
$(M_{i}^{ })_{max}$ , $(\Gamma_{i}^{ee})_{max}$ (Table 1).

The difference between the right-hand and left-hand sides of (24) is minimal
at masses $M_{i}^{Best}$ and electronic widths $\Gamma_{i}^{Best}$
and less than $3\%$ for the ${\cal J}$/$\Psi$  family for all positive $Q^{2}$
and less than $9\%$ for the $\Upsilon$  family for all $Q^{2}$ from interval
$-4 m_{b}^{2} + 10 \GeV^{2} < Q^{2} < \infty$.
$(M_{i}^{ })_{min}$ , $(\Gamma_{i}^{ee})_{min}$ ,
$(M_{i}^{ })_{max}$ , $(\Gamma_{i}^{ee})_{max}$ are minimal (maximal) value
of masses and electronic widths of mesons such that the right-hand and
left-hand sides of (24) differ by less than $20\%$ if $Q^{2}$ is in the
interval where the O.E. is valid.

\section{Acknowledgment}

The authors are grateful to Frank Paige for his review of the questions
concerned and appreciate his comments.  We also thank the
staff of the SSC Computer Center.  We appreciate the efforts of Linda Fowler
and thank her for her contribution in preparation of this article for
publication.

\section{References}
{\frenchspacing
\begin{tabbing}
1.~~~\=M.A.~Shifman,